\documentclass{osa-article}

\newcommand{\req}[1]{Eq.~(\ref{#1})}
\newcommand{\rfig}[1]{Fig.~(\ref{#1})}

\newcommand{\change}[1]{\textcolor{black}{#1}}

\newcommand{\rem}[1]{\phantom{#1}}

\usepackage[nottoc,numbib]{tocbibind}

\definecolor{MyDarkBlue}{rgb}{0.2,0.2,0.9}

\usepackage{hyperref}
\hypersetup{
    colorlinks=true,
    linkcolor=MyDarkBlue, 
    urlcolor=MyDarkBlue,
    citecolor=MyDarkBlue
}

\usepackage{subfig}
\usepackage{soul}
\usepackage{xcolor}
\journal{osajournal}

\articletype{Research Article}

\begin{document}

\title{Analytical Approximation of the Second-Harmonic Conversion Efficiency}

\author{
John R.~Daniel,\authormark{1}
Shan-Wen Tsai,\authormark{1} 
and 
Boerge Hemmerling\authormark{1}
}

\address{
\authormark{1}Department of Physics and Astronomy, University of California, Riverside, USA\\
}

\email{\authormark{*}jdani017@ucr.edu} 

\homepage{http://molecules.ucr.edu} 

\begin{abstract}
The second-harmonic generation process of a focused laser beam inside a nonlinear crystal is described by the Boyd-Kleinman theory. Calculating the actual conversion efficiency and upconverted power requires the solution of a double integral that is analytically intractable. 
We provide an expression that predicts the exact gain coefficient within an error margin of less than 2\% over several orders of magnitude of the confocal parameter and as a function of the walk-off parameter. Our result allows for readily tuning the beam parameters to optimize the performance of the upconversion process and improve optical system designs.
\end{abstract}

\section{Introduction}
As research on quantum physics continuously expands to a wider range of atomic and molecular systems, it becomes more important for probing such systems to generate laser frequencies in regimes that are not accessible directly with the available robust solid-state laser technology \cite{Weber1999}.
In particular, the ultra-violet regime and large ranges around 400 to 600\,nm are affected by the lack of readily usable laser sources. On the other hand, there is an ever growing demand to cover this regime since many applications, including cooling and trapping of molecules with ultraviolet transitions \cite{DiRosa2013}, laser cooling of ions \cite{Eschner2003,Leibfried2003}, optical clocks \cite{Brewer2019,Ludlow2015}, and precision measurements \cite{Beyer2017}, require high-power continuous-wave laser sources at short wavelengths.
To mitigate this technological challenge, a typical approach starts with a high-power source at a longer wavelength that is then frequency-upconverted inside a crystal whose polarization depends non-linearly on the electric field to generate the desired light at the shorter wavelength \cite{Shukla2014,Devi2016,Jurdik2002}.

Here, we focus on the second-harmonic generation (SHG) process where the conversion efficiency scales quadratically with the electric field. As a result, in applications that employ pulsed lasers, the short, intense electric field already provides a sufficiently high conversion efficiency in a single-pass crystal configuration \cite{Ghosh2019}. 
On the other hand, continuous-wave applications often require a resonant enhancement cavity that matches the optimal beam waist inside the crystal to provide high upconverted power \cite{Hannig2018}.
One of the key points in designing these optical systems is to determine the optimal laser beam parameters for a given crystal to allow for maximum conversion efficiency. These parameters include the beam shape, optical geometry, and particular choices regarding the dimensions, orientation, and chemical composition of the nonlinear crystal used for upconversion \cite{Alford2001,Sasaki2000}.

In their paper \cite{Boyd1968,Kleinman1966}, Boyd and Kleinman calculated the conversion efficiency as a function of these parameters and derived a theoretical expression in the form of a double integral that is analytically unsolvable. 
Thus, obtaining a simple approximation is useful for the expedient optimization of system parameters and to study general tendencies while varying system parameters.
Here, we propose an empirical analytical expression that reproduces the Boyd-Kleinman conversion efficiency factor over several orders of magnitude for the focal parameter with an error of <2\%. This constitutes an improvement over previous approximations by almost two orders of magnitude \cite{Chen2003,Ramazza2002}.

In the following, we briefly outline the results of the Boyd-Kleinman derivation and show that the general result can be dimensionally reduced by a simple coordinate rotation. We then discuss our proposed approximation and provide an expression that recovers the value of the Boyd-Kleinman integral as a function of the walk-off and the focal parameters. Finally, we provide an approximate expression for the optimal maximum value of the conversion efficiency.

\section{Conversion Efficiency of Second-Harmonic Generation}

The Boyd-Kleinman (BK) SHG theory \cite{Boyd1968,Kleinman1966} describes the efficiency at which a Gaussian laser beam with frequency $\omega_1$, focused into a uniaxial nonlinear crystal, is upconverted to a frequency $\omega_2 = 2 \omega_1$.

Using the derivation of Boyd and Kleinman and assuming that the laser beam is focused in the center of the crystal and absorption losses are negligible, the power of the upconverted wave in SI units can be expressed as \cite{Boyd1968,Kleinman1966}
\begin{eqnarray}\label{eq:powers}
P_\textrm{SHG} = 
   \left(
   \frac{2 \omega_1^2 d_\textrm{eff}^2 L_c k_1}{\pi \epsilon_0 c^3 n_1^2 n_2} 
   \cdot
   P_1^2
   \right)
   \cdot h_\textrm{SHG}(\sigma, \xi, B)\, ,
\end{eqnarray}
where $P_1$ is the power of the fundamental wave,
$d_\textrm{eff}$ is the effective non-linear coefficient, $n_1$ and $n_2$ are the respective indices of refraction for the fundamental and second-harmonic laser beams, $L_c$ is the crystal length, and $k_1$ is the fundamental wavenumber. $\epsilon_0$ and $c$ are the vacuum permittivity and vacuum speed of light. The function $h_\textrm{SHG}(\sigma,\xi,B)$ is given by \cite{Boyd1968,Kleinman1966}
\begin{eqnarray}\label{eq:2dint}
h_\textrm{SHG}(\sigma,\xi,B) &=& 
\frac{1}{4\xi} 
\int\displaylimits_{-\xi}^{+\xi}\!\!d\tau_1 
\int\displaylimits_{-\xi}^{+\xi}\!\!d\tau_2
\frac{e^{\imath \sigma \cdot (\tau_1 - \tau_2)} 
\cdot 
e^{-\frac{B^2 ( \tau_1 - \tau_2 )^2}{\xi}}}{
(1 + \imath \tau_1) \cdot
(1 - \imath \tau_2)}\, ,
\end{eqnarray}
\change{
with the usual definitions in the literature, $\sigma = b\Delta k/2$, the walk-off parameter $B = (\rho \sqrt{L_c k_1})/2$.
The parameter $\rho$ is the walk-off angle of the crystal, $\Delta k = 2k_1 - k_2$ is the phase mismatch, and $b = \varpi_0^2 k_1$ is the confocal parameter, for which $\varpi_0$ is the waist of the Gaussian beam. 
The parameter $\xi = L_c/b$ corresponds to the ratio of the crystal length and the confocal parameter.
The case of $\xi \ll 1$ corresponds to a loose focus and $\xi \gg 1$ corresponds to a tight focus.} Both the walk-off angle and the effective non-linear coefficient are material parameters and are tabulated in the literature \cite{Dmitriev1999,Smith2018}. \change{A few examples of commonly used crystals are Potassium Titanyl Phosphate (KTP), Lithium Triborate (LBO), and Barium Borate (BBO). For $1 \mu \textrm{m}$ laser light in a $L_c = 1 \textrm{cm}$ crystal, KTP has $B \approx 0.5$, LBO has $B \approx 0.9$, and BBO has $B \approx 7$. For this crystal size and laser wavelength, the range of $0.001 < \xi < 100$ covers focus waists in the range of $4 \mu\textrm{m} < \varpi_0 < 1200 \mu\textrm{m}$.}

To maximize the power of the second-harmonic wave, the BK integral (\req{eq:2dint}) needs to be maximized by tuning the two free parameters, the waist of the fundamental beam and the phase mismatch parameter. The latter is adjusted by, for example, changing the temperature of the crystal or the angle between the incident beam and the crystal axis to vary the refractive indices of fundamental and harmonic wave. The beam waist is chosen to be close to the optimal waist inside the crystal by mode-shaping the incident beam.

Due to the similarity to a Voigt function, \req{eq:2dint} cannot be integrated and optimized analytically. As a first simplification, we note that the two-dimensional integral can be simplified by utilizing the coordinate transformation
\begin{eqnarray*}
\tau_1 = \frac{x+y}{\sqrt{2}}
\quad,\quad
\tau_2 = \frac{-x+y}{\sqrt{2}} \quad.
\end{eqnarray*}
This $\pi/4$-rotation allows for analytical integration along the $y$-direction, and yields after simplification
\begin{eqnarray}
\label{eq:1dint}
&&h_\textrm{SHG}(\sigma, \xi, B) =
-\frac{2}{\xi}\cdot \Re\left[
\int\displaylimits_{0}^{\sqrt{2} \xi}\!\!dx
\frac{
e^{\imath \sqrt{2} \sigma x}
\cdot
e^{-\frac{2 B^2 x^2}{\xi}}
\cdot
\arctan \left(
\frac{x-\sqrt{2} \xi}{
\sqrt{2} + \imath x
}
   \right)}
   {\sqrt{2} + \imath x}
   \right].
\end{eqnarray}
Here, $\Re[]$ signifies the real part of the integral value.
This reduction of the integral is computationally simpler, however, still analytically unsolvable for the same reasons. In the following, we describe an analytical approximation that readily provides precise values to the BK integral.

\begin{figure}[ht]
    \centering
\includegraphics[scale=0.7]{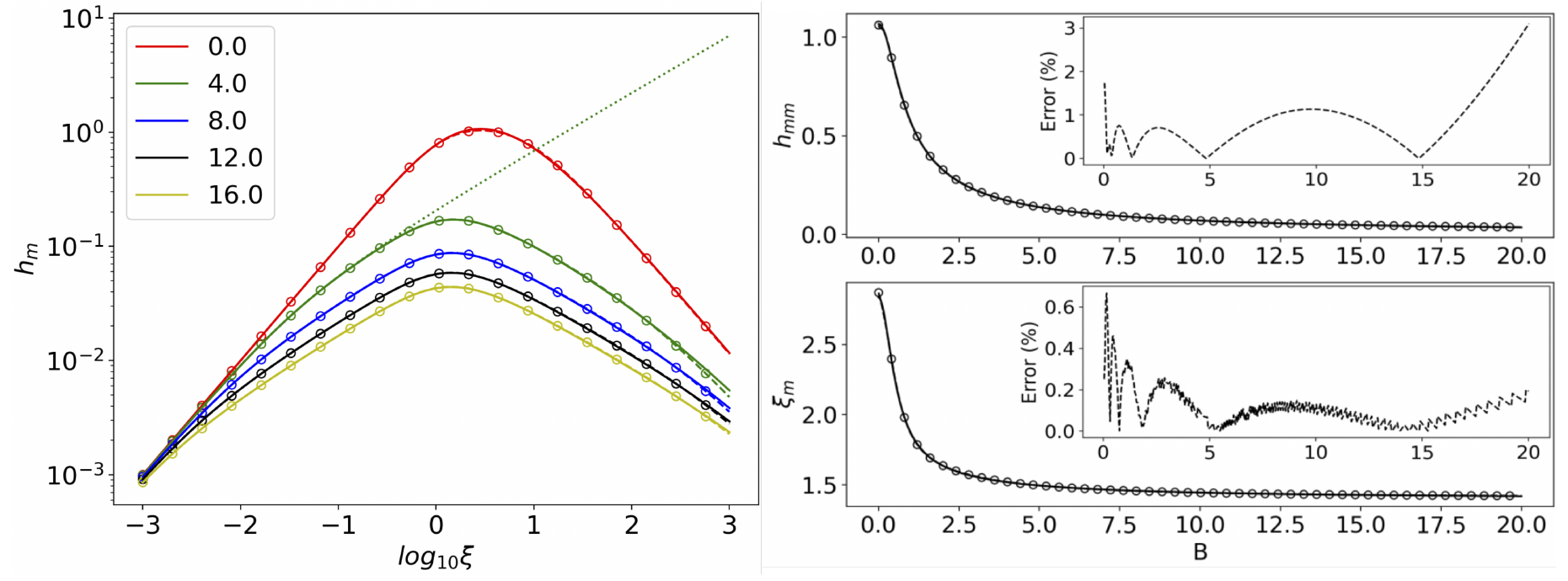}
\caption{
\label{fig:hmvxi}
   {\bf Left:} Comparison of values obtained from (\ref{eq:1dint}) via numerical integration (solid) and the approximate values obtained from (\ref{eq:hmshg}) (dashed with circles) for various values of $B$. The circles are added as a guide to the eye since the approximation and the full integration are virtually identical for all $\xi \lessapprox 100$. The additional line (dotted) for $B = 4$ is a plot of the small-$\xi$ approximation in \req{eq:hm_small_xi}.
   {\bf Right:} Plotted are $h_\textrm{mm}(B)$ and $\xi_\textrm{m}(B)$ generated from the numerical solution to \req{eq:1dint} (solid) and the approximate values generated from \req{eq:hmm} and \req{eq:xim} (dashed with circles). The percent difference between the numerical and approximate is also shown (inset).}
\end{figure}

\section{Approximation of the Boyd-Kleinman Integral}

We define the maximum of the BK integral (\req{eq:2dint} or \req{eq:1dint}) with respect to the mismatch parameter $\sigma$ as 
\begin{eqnarray}
h_\textrm{m,SHG}(\xi,B) = \textrm{max}[h_\textrm{SHG}(\sigma,\xi,B)]_\sigma\,.
\end{eqnarray}
To construct an approximation to the BK integral, we first integrate $h_{m, \textrm{SHG}}$ for small $\xi$. For small $\xi$, the maximum value of $h_{m, \textrm{SHG}}$ becomes independent of $\sigma$ and the Gaussian term becomes the dominant contribution to the integral. This approximation is analytically integrable for $\xi \ll 1$, yielding
\begin{eqnarray}
\label{eq:hm_small_xi}
   &&h_\textrm{m, SHG} (\xi, B) \approx 
\kappa
   \cdot \xi\,\quad
   ;\quad
   \kappa = 
   \frac{
e^{-\delta^2} - 1 + \sqrt{\pi} \cdot \delta \cdot \textrm{erf}(\delta)
}
   {\delta^2}
\end{eqnarray}
with $\delta = 2 B \sqrt{\xi}$. This approximation exactly reproduces the full BK integral for values of $\xi < 0.5$. An example for $(B=4)$ is shown in \rfig{fig:hmvxi}, along with the exact numerical integration. Note that \req{eq:hm_small_xi} also recovers the case of zero walk-off $(B=0)$, as $\kappa \rightarrow 1$ for $\xi \rightarrow 0$.

Based on this small $\xi$ expansion, we propose the following
analytical function to approximate the BK integral for a wide range of $\xi$ values as
\begin{eqnarray}\label{eq:hmshg}
   h_\textrm{m, SHG} (\xi, B)
\approx
\frac{
\arctan
\left(
c_1 \cdot \kappa \cdot \xi
\right)
}
{
c_1 + c_2 \cdot \xi \cdot \arctan(c_3 \cdot \xi)
}
\quad
\textrm{for}\,\,
(0 < \xi \lessapprox 100)
\end{eqnarray}
This function is constructed such that it converges to \req{eq:hm_small_xi} for small $\xi$ and is proportional to $1/\xi$ for $\xi \rightarrow \infty$, recovering the limiting cases of the BK integral.

\begin{figure}
    \centering
    \includegraphics[scale=0.6]{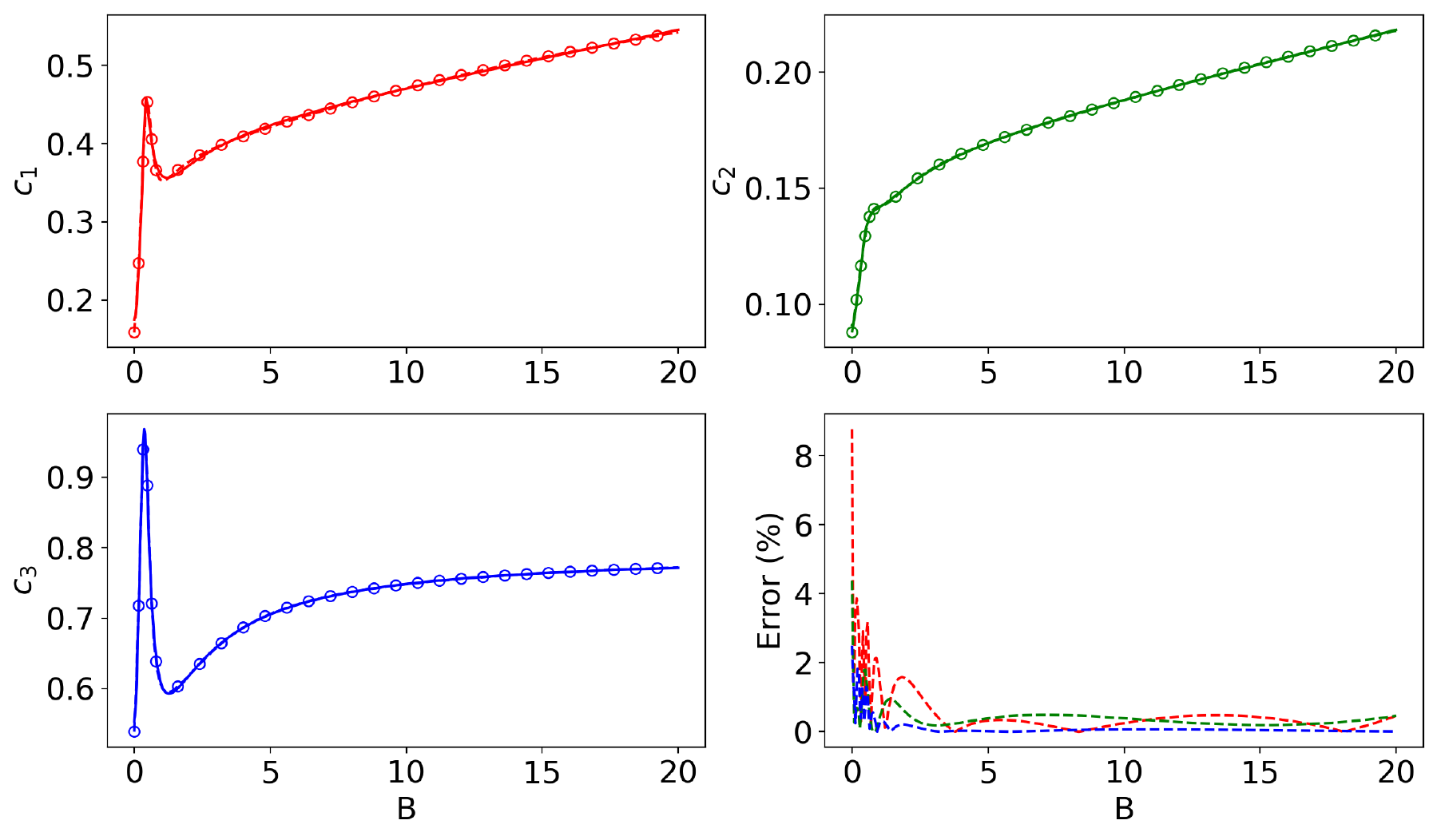}
    \caption{
    The parameters $c_1$, $c_2$, and $c_3$ are plotted as a function of the walk-off parameter $B$, showing their numerical solution (solid) from a least-squares fit of \req{eq:hmshg} and the approximate value (dashed with circles) generated from \req{eq:ceq1}--\req{eq:ceq3}. The percent difference between the numerical and approximate values is shown in the bottom right graph following the same color coding.
    \label{fig:params}
    }
\end{figure}

The parameters $c_1$, $c_2$, and $c_3$ are relatively smooth functions of the walk-off parameter $B$, as is shown in \rfig{fig:params}, and can be approximated by polynomial expressions. We find very good agreement with the BK integral with the form
\begin{eqnarray}\label{eq:ceq1}
c_{1}(B) &=& 0.876 - \frac{18.8}{B + 36.5} + \frac{0.0166}{0.0693 + (B - 0.440)^2} - \frac{0.283}{0.931 + (B + 0.516)^3}\\
\label{eq:ceq2}
c_{2}(B) &=& 0.530 - \frac{36.0}{B + 95.1} + \frac{0.0103}{0.332 + (B - 0.569)^2} - \frac{0.497}{4.69 + (B + 1.15)^3}\\
\label{eq:ceq3}
c_{3}(B) &=& 0.796 - \frac{0.506}{B + 0.378} + \frac{0.0601}{0.421 + (B - 0.673)^2} + \frac{0.0329}{0.0425 + (B - 0.221)^3}
\end{eqnarray}

In \rfig{fig:hmvxi} (left) we plot $h_\textrm{m}$ for a range of magnitudes of $\xi$ for various values of $B$. Due to the incorporation of the small $\xi$ expansion, our proposed approximation is virtually identical to the numerical integration except for $\xi > 100$.

\begin{figure}[ht]
\centering
\hspace*{-1.0cm}
\includegraphics[scale=0.6]{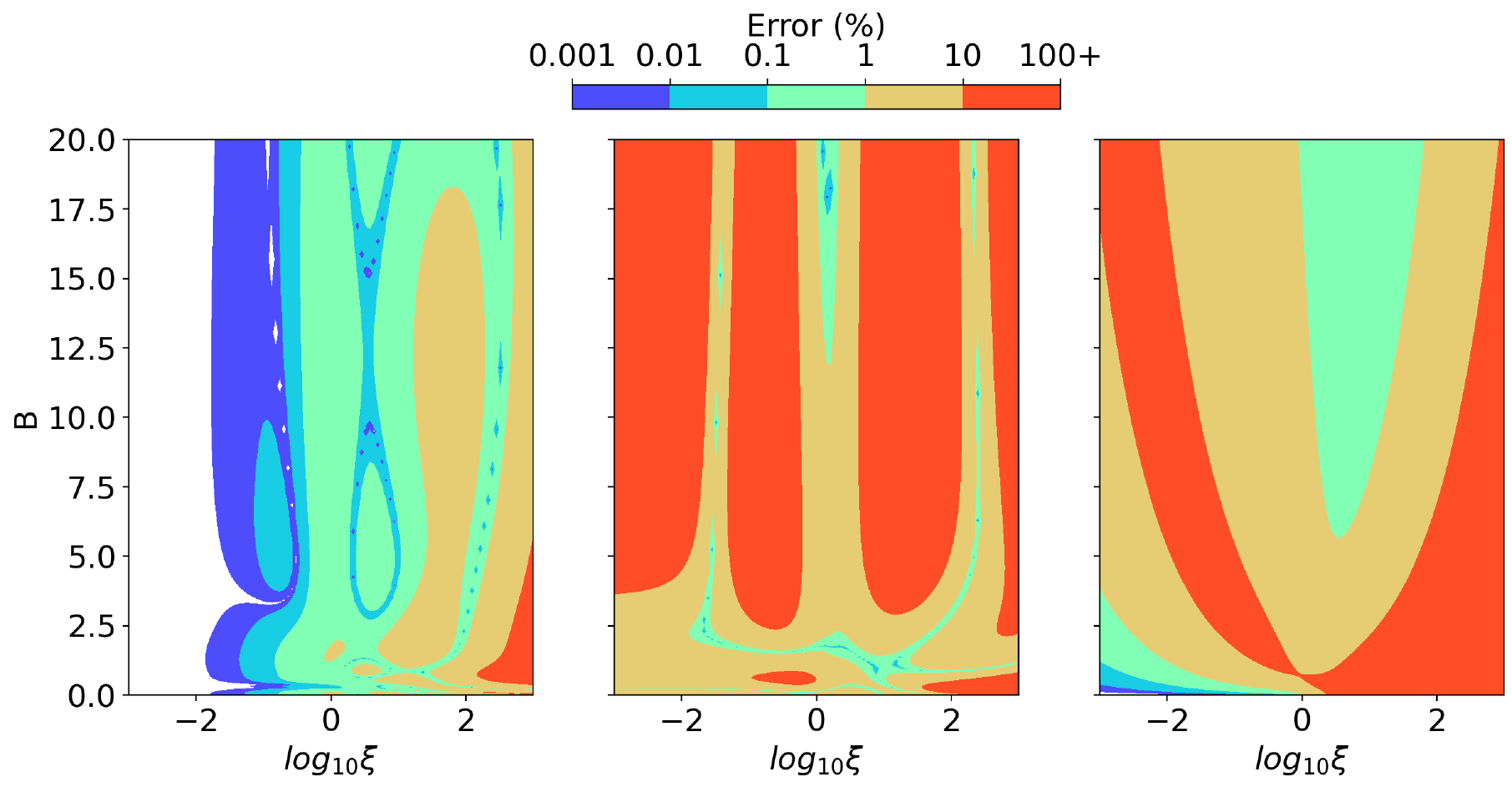}
\caption{
\label{fig:resids}
   Percent difference between numerical integration of the BK integral \req{eq:1dint} and our work \req{eq:hmshg} (left) and previous work (middle \cite{Chen2003} and right \cite{Ramazza2002}). Red regions correspond to errors greater than 10\%, orange to the range of 1\% to 10\%, etc. The white region on the left corresponds to errors smaller than $0.001$\%.
}
\end{figure}

Furthermore, we compare in \rfig{fig:resids} the precision of our approximation with previously proposed expressions \cite{Ramazza2002,Chen2003}. 
Our approximation of $h_\textrm{m}$ greatly improves the range of values of $\xi$ up to $\approx 100$ for which the approximation differs by less than 2\% from the direct numerical integration. Previous work accurately resembles the BK integral in regions around the cusp or is limited to smaller ranges of $\xi$ whereas outside these regions the residual errors grow considerably, as can be seen in the \rfig{fig:resids} (middle and right).

Finally, we also reproduce the optimal value $\xi_\textrm{m}$, \rem{i.e.~}\change{from which we can calculate} the beam waist inside the crystal that maximizes the SHG output, and the corresponding conversion efficiency $h_\textrm{mm}(B) = h_\textrm{m}(\xi_\textrm{m},B)$ as a function of the walk-off parameter $B$.
The values of $h_\textrm{mm}$ and $\xi_\textrm{m}$ can be approximated as
\begin{eqnarray}
\label{eq:hmm}
h_\textrm{mm}(B) &=& 0.01459 + \frac{20.88}{(B + 13.69)^2 - 181.0} - \frac{0.2954}{(B + 0.5840)^4 + 0.01749}\\
\label{eq:xim}
\xi_\textrm{m}(B) &=& 1.410 + \frac{5.924}{(B + 3.508)^2 - 1.762} + \frac{0.7232}{(B + 0.4640)^4 + 0.7577}.
\end{eqnarray}

In \rfig{fig:hmvxi} (right top and bottom) we compare the values for $h_\textrm{mm}$ and $\xi_\textrm{m}$ that are obtained from the approximation in \req{eq:hmm}--\req{eq:xim} with their counterparts obtained from the direct numerical integration. Here, our approximation reproduces $h_\textrm{mm}$ within a margin of $\approx 3$\% and $\xi_\textrm{m}$ within a margin of $< 1$\% both for $B \le 20$.

\change{In summary, the conversion efficiency of a focused Gaussian laser beam in a second-harmonic generation process inside a non-linear crystal with walk-off parameter $B$ can be readily determined by the following steps:
Given the crystal length and the confocal parameter, one can calculate $\xi$ and determine the expected SHG conversion efficiency $h_\textrm{m,SHG}(\xi, B)$ from \req{eq:hmshg}, while plugging in the parameters $c_{1,2,3}(B)$ (\req{eq:ceq1}-\req{eq:ceq3}). The maximum possible conversion efficiency $h_\textrm{mm}(B)$ for this crystal configuration is then given by \req{eq:hmm}. The corresponding optimal waist of the Gaussian beam is then given by
$\varpi_0 = \sqrt{L_c/k_1 \xi_\textrm{m}(B)}$,
where $\xi_\textrm{m}(B)$ is given by \req{eq:xim}.
}

\section{Conclusion}

In this manuscript, we provide two simplifications to the two-dimensional BK integral to calculate the conversion efficiency of a focused Gaussian laser beam inside a uniaxial crystal. First, we reduce the dimension of the integral by rotating the coordinate system and allowing for exact analytical integration along one direction. Second, we approximate the resulting expression with a simple analytical function with three parameters as a function of the walk-off parameter. 
We show that these parameters can be approximated with inverse polynomials with which the conversion efficiency for SHG can be predicted with an error margin of $\approx 2$\% over a large parameter space for values of $\xi \lessapprox 100$.
\change{Finally, our treatment may be applicable to the more general case of systems that employ sum-frequency generation to create up- or down-converted light.}
In this case, the additional laser beam adds a free parameter to the system that results in larger computational requirements \cite{Guha1980} and analytical progress along the lines of the work presented here would be highly beneficial. We leave this analysis for a future investigation.

\section*{Funding}
We acknowledge funding from the National Science Foundation (NSF) (1839153).

\section*{Disclosures}

The authors declare no conflicts of interest.

\bibliography{references}

\end{document}